\documentclass[preprint,showpacs,preprintnumbers,amsmath,amssymb,superscriptaddress]{revtex4}
\usepackage{graphicx,color}
\usepackage{amssymb}
\usepackage{amsmath}
\usepackage{latexsym}
\usepackage{amsxtra}
\usepackage{amscd}
\usepackage{dcolumn}
\usepackage{bm}

\usepackage{hyperref}

\begin{document}

%\preprint{APS/123-QED}

\title{Discovery of the shape coexisting $0^+$ state in $^{32}$Mg by a two neutron transfer reaction}% Force line breaks with \\

\author{K.~Wimmer}
\author{T.~Kr\"oll}\altaffiliation[present address ]{Institut f\"ur Kernphysik, Technische Universit\"at Darmstadt, Germany}
\author{R.~Kr\"ucken}
\author{V.~Bildstein}
\author{R.~Gernh\"auser}\affiliation{Physik Department E12, Technische Universit\"at M\"unchen, 85748 Garching, Germany}
\author{B.~Bastin}
\author{N.~Bree}
\author{J.~Diriken}
\author{P.~Van Duppen}
\author{M.~Huyse}
\author{N.~Patronis}\altaffiliation[present address ]{Department of Physics, The University of Ioannina, GR-45110 Ioannina,
Greece}
\author{P.~Vermaelen}\affiliation{Instituut voor Kern- en Stralingsfysica, Katholieke Universiteit Leuven, Belgium}
\author{D.~Voulot}
\author{J.~Van de Walle}
\author{F.~Wenander}\affiliation{CERN, Gen\`eve, Switzerland}
\author{L.M.~Fraile}\affiliation{Dpto.~de F\'\i sica At\'omica, Molecular y Nuclear, Universidad Complutense, Madrid, Spain}
\author{R.~Chapman}
\author{B.~Hadinia}
\author{R.~Orlandi}
\author{J.F.~Smith}\affiliation{School of Engineering and Science, Univ.~of the West of Scotland, Paisley, Scotland, UK}
\author{R.~Lutter}
\author{P.G.~Thirolf}\affiliation{Fakult\"at f\"ur  Physik, Ludwig-Maximilians-Universit\"at M\"unchen, Garching, Germany}
\author{M.~Labiche}\affiliation{Daresbury Laboratory, Warrington, UK}
\author{A.~Blazhev}
\author{M.~Kalk\"uhler}
\author{P.~Reiter}
\author{M.~Seidlitz}
\author{N.~Warr}\affiliation{Institut f\"ur Kernphysik, Universit\"at zu K\"oln, Germany}
\author{A.O.~Macchiavelli}
\author{H.B.~Jeppesen}\affiliation{Lawrence Berkeley National Laboratory, USA}
\author{E.~Fiori}
\author{G.~Georgiev}\affiliation{CSNSM-IN2P3-CNRS, Universit\'e Paris-Sud 11, 91405 Orsay, France}%Centre de Spectrom\'etrie Nucl\'eaire et de Spectrom\'etrie de Masse, Orsay, France}
\author{G.~Schrieder}\affiliation{Institut f\"ur Kernphysik, Technische Universit\"at Darmstadt, Germany}
\author{S.~Das~Gupta}
\author{G.~Lo~Bianco}
\author{S.~Nardelli}\affiliation{Dipartimento di Fisica, Universit\`a di Camerino, Italy}
\author{J.~Butterworth}\affiliation{Nuclear Physics Group, Department of Physics, University of York, UK}
\author{J.~Johansen}
\author{K.~Riisager}\affiliation{University of Aarhus, Denmark}
%\author{T.~Davinson}\affiliation{University of Edinburgh, Scotland, UK}


\begin{abstract}
The {\it Island of Inversion} nucleus $^{32}$Mg has been studied by a (t,p) two neutron transfer reaction in inverse kinematics at REX-ISOLDE. The shape coexistent excited $0^+$ state in $^{32}$Mg has been identified by the characteristic angular distribution of the protons of the $\Delta L =0$ transfer. The excitation energy of 1058~keV is much lower than predicted by any theoretical model. The low $\gamma$-ray intensity observed for the decay of this $0^+$ state indicates a lifetime of more than 10 ns. Deduced spectroscopic amplitudes are compared with occupation numbers from shell model calculations.
\end{abstract}

\pacs{
21.10.Hw 	%Spin, parity, and isobaric spin
25.40.Hs 	%Transfer reactions
27.30.+t 	%20 < A < 38
29.38.-c 	%Radioactive beams
}% PACS, the Physics and Astronomy
\maketitle
The evolution of shell structure in exotic nuclei as a function of proton ($Z$) and neutron ($N$) number is currently at the center of many theoretical and experimental investigations~\cite{sorlin08,kruecken10}.
It has been realized that the interaction of the last valence protons and neutrons, in particular the monopole component of the residual interaction between those nucleons, can lead to significant shifts in the single-particle energies, leading to the disappearance of classic shell closures and the appearance of new shell gaps~\cite{otsuka01}. A prominent example is the collapse of the $N=20$ shell gap in the neutron-rich oxygen isotopes where instead a new magic shell gap appears for $^{24}$O at $N=16$~\cite{hoffman08,kanungo09}.
Recent work showed that the disappearance of the $N=20$ shell can be attributed to the monopole effect of the tensor force~\cite{otsuka01,otsuka05,otsuka10}. The reduced strength of the attractive interaction between the proton $\pi d_{5/2}$ and the neutron $\nu d_{3/2}$ orbitals causes the $\nu d_{3/2}$ orbital to rise in energy and come closer to the $\nu f_{7/2}$ orbital.
In regions without pronounced shell closures correlations between the valence nucleons may become as large as the spacing of the single particle energies. This can thus lead to particle-hole excitations to higher-lying single-particle states enabling deformed configurations to be lowered in energy. This may result in low-lying collective excitations, the coexistence of different shapes at low energies or even the deformation of the ground state for nuclei with the conventional magic number $N=20$.

Such an effect occurs in the {\it Island of Inversion}, one of most studied regions of exotic nuclei in the nuclear chart. In this region of neutron-rich nuclei around the magic number $N=20$ strongly deformed ground states in Ne, Na, and Mg isotopes have been observed~\cite{thibault75,detraz79,guillemaud84,doornenbal09}. Due to the reduction of the $N=20$ shell gap, quadrupole correlations can enable low-lying  deformed $2p-2h$ intruder states from the $fp$-shell to compete with spherical normal neutron $0p-0h$ states of the $sd$-shell. In this situation the promotion of a neutron pair across the $N=20$ gap can result in deformed intruder ground states. Consequentially the competition of two configurations can lead to the coexistence of spherical and deformed $0^+$ states in the neutron rich $^{30,32}$Mg nuclei~\cite{heyde91}.

Coulomb excitation experiments have shown that $^{30}$Mg has a rather small $B(\text{E2})$ value for the $0^+_\text{gs}\rightarrow 2^+_1$ transition~\cite{niedermaier05,pritychenko99} placing this nucleus outside the {\it Island of Inversion}. The excited deformed $0^+$ state in $^{30}$Mg at 1789~keV was recently identified at ISOLDE by its E0 decay to the ground state~\cite{mach05,schwerdtfeger09}. The small electric monopole strength $\rho^2(\text{E0};\,0^+_2 \rightarrow 0^+_\text{gs})$ is consistent with a small mixing amplitude between the two shape coexisting $0^+$ states. Beyond-mean-field (BMF) calculations, which incorporate configuration mixing using the finite-range density-dependent Gogny force with the D1S parametrization~\cite{berger84}, have reasonably reproduced this coexistence scenario. They predict an excitation energy of 2.11 MeV for the $0^+_2$ state in $^{30}$Mg and only weak mixing between the two $0^+$ states. In $^{31}$Mg a recent measurement of the ground state spin $J^\pi = 1/2^+$~\cite{neyens05} could only be explained by a dominant intruder configuration in the ground state~\cite{marechal05}, thus placing $^{31}$Mg exactly on the border of the {\it Island of Inversion}.

The large $B(\text{E2};\,0^+_\text{gs} \rightarrow 2^+_1)$ value~\cite{motobayashi95,pritychenko99} for $^{32}$Mg has clearly established its strongly deformed ground state. However, while some spectroscopic information on excited states in $^{32}$Mg is available from $\beta$-decay studies~\cite{mach05,mattoon07} and reactions~\cite{gelin_phd}, no excited $0^+$ state has been observed so far. Shell model (SM) calculations that correctly describe the deformed ground state in $^{32}$Mg  predict the spherical excited $0^+$ state at 1.4~MeV~\cite{caurier01} and 3.1 MeV~\cite{otsuka04}. The BMF calculations in Ref.~\cite{rodriguez02} predict a spherical shape coexisting $0^+$ state at about 1.7~MeV. The energy of this state may be sensitive to the strength of the quadrupole correlations as well as the single-particle energies and cross-shell mixing. Therefore, the observation of this shape coexisting spherical excited $0^+$ state in $^{32}$Mg would provide important input to refine the theoretical description of the transition into the {\it Island of Inversion}.

It is the purpose of this letter to report on the first experimental observation of the long-sought excited shape coexisting $0^+$ state in $^{32}$Mg by a two neutron transfer (t,p) reaction ($Q=-295(20)$~keV). In this work a radioactive tritium target was used for the first time in combination with a radioactive heavy ion beam. The shape of the angular distribution of protons allows to unambiguously determine the angular momentum transfer $\Delta L$ of the reaction. Results on excitation energies and spectroscopic amplitudes extracted from the cross section are compared with shell model calculations.

The experiment was performed at REX-ISOLDE (CERN) where the $^{30}$Mg beam was produced by the 1.4 GeV proton beam with an intensity of $3\cdot 10^{13}$ protons/pulse from the CERN PS Booster impinging on an UC$_\text{x}$/graphite target. Mg atoms were selectively ionized using the RILIS laser ion source, the ions were mass separated in the ISOLDE General Purpose Separator and post-accelerated to 1.8~MeV/u with the REX-ISOLDE linear accelerator after bunching and charge breeding~\cite{kester03}. The $4.6(5)\cdot 10^4$~part/s beam was sent to the T-REX~\cite{bildstein07} transfer setup surrounded by the MINIBALL $\gamma$-ray detector array~\cite{eberth01}. T-REX consists of position sensitive $\Delta E-E$ telescopes allowing for the detection and identification of the light target-like reaction products. The solid angle coverage of T-REX amounts to 58.5\% of $4\pi$. The target in the center of the setup is based on a thin strip of 500~$\mu$g/cm$^2$ thick metallic Ti foil loaded with an atomic ratio $^{3}$H/Ti of 1.5 corresponding to a target thickness of 40~$\mu$g/cm$^2$ $^{3}$H. The activity of the target was 10~GBq.

\begin{figure}[h]
\includegraphics[width=0.48\textwidth]{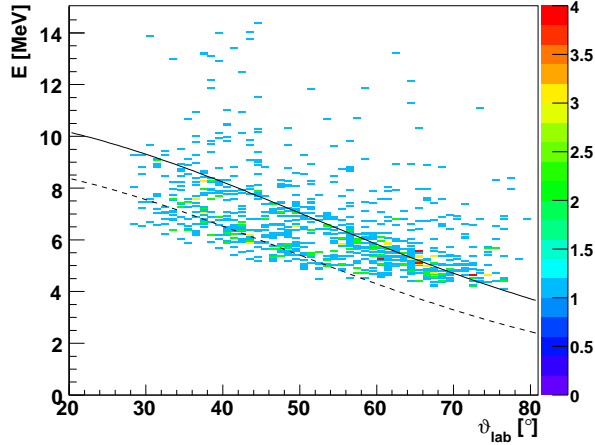}
\caption{Energy versus $\vartheta_\text{lab}$ spectrum for $^{30}$Mg on the tritium target for unambiguously identified protons. The two lines show the kinematic curves for the ground state (solid) and an excited state at 1.1~MeV (dashed) (color online).}
\label{fig:eth}
\end{figure}

\begin{figure}[h]
\includegraphics[width=0.48\textwidth]{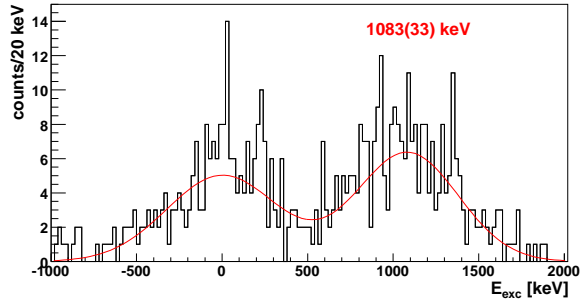}
\caption{Excitation energy of the ejectile $^{32}$Mg reconstructed from the measured energy of protons in T-REX. The excitation energy of the observed state amounts to $1083(33)$~keV.}
\label{fig:exfit}
\end{figure}

After identifying protons with the $\Delta E-E$ telescopes, for each event the excitation energy of the ejectile nucleus $^{32}$Mg can be determined from the energy and emission angle of the recoiling proton. Fig.~\ref{fig:eth}
shows the proton energy vs. the laboratory scattering angle for all clearly identified protons, while Fig.~\ref{fig:exfit} shows the reconstructed excitation energy spectrum of $^{32}$Mg for protons in the angular range where the full excitation energy spectrum can be observed.
Two states have been observed, the ground state of $^{32}$Mg and an excited state at $1083(33)$~keV.
The resolution is limited by the energy loss of the beam in the target, and of the recoiling protons in the target and protection foils in front of the T-REX detectors.
Contributions from compound nucleus reactions can be neglected, as the neutron rich products will mostly evaporate neutrons due to their lower separation energy. In order to avoid fusion reactions with the Ti carrier material of the target the beam energy was limited to 1.8~MeV/u.

\begin{figure}[h]
\includegraphics[width=0.48\textwidth]{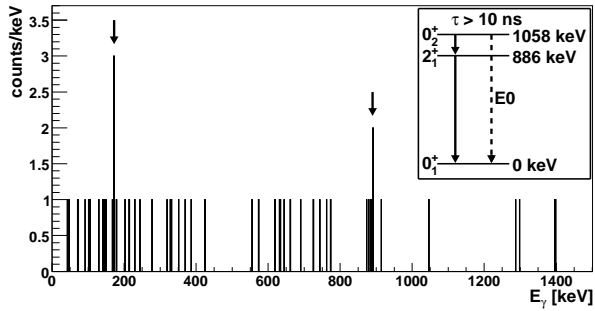}
\caption{$\gamma$-ray energy spectrum in coincidence with protons from transfer to the excited state. The inset shows the suggested partial level scheme of $^{32}$Mg.}
\label{fig:gamma}
\end{figure}

The $\gamma$-ray energy spectrum recorded in coincidence with protons from the (t,p) reaction to the excited state is shown in Fig.~\ref{fig:gamma}.
The two peaks at 172~keV ($6(3)$ counts) and 886~keV ($4(2)$ counts) suggest a cascade from a level at 1058~keV through the well known $2^+$ level at 886 keV. This low count rate further confirms that the $2^+$ state has not been populated directly in the (t,p) reaction. The value of 1058~keV for the excitation energy of the initial state is consistent with the observed excitation energy reconstructed from the proton energies.

\begin{figure}[h]
\includegraphics[width=0.48\textwidth]{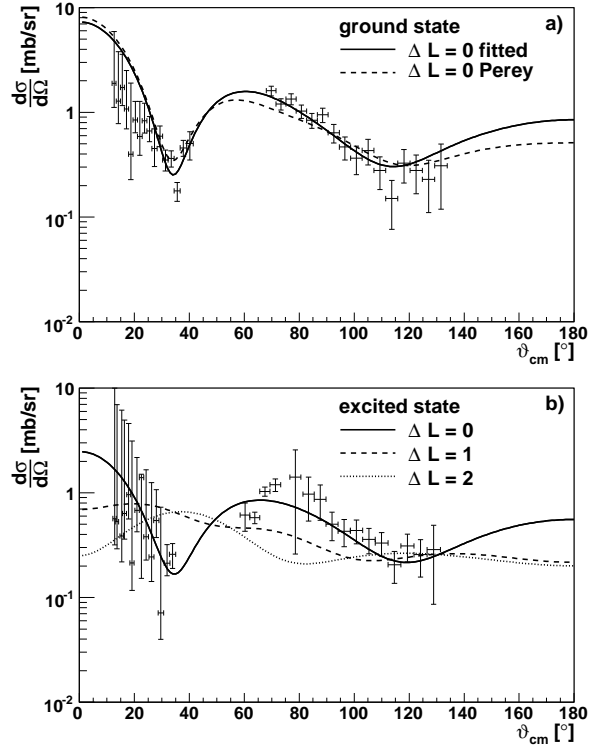}
\caption{Proton angular distributions for the reaction t$(^{30}$Mg,p)$^32$Mg populating th ground state (a) and the 1058 keV state (b). The curves in panel a) are from DWBA calculations with fitted (solid) and global (dashed) optical model parameters~\cite{perey76}. Panel b) includes DWBA calculations for $\Delta L =0$ (solid), $\Delta L=1$ (dashed) and $\Delta L=2$ (dotted) transfer.
The angle range around $\vartheta_\text{cm}=50^\circ$ corresponds to $\vartheta_\text{lab}=90^\circ$ which is not covered by the T-REX detector.}
\label{fig:angfit}
\end{figure}

The angular momentum of both states can be determined from the proton angular distributions, which are shown in Fig.~\ref{fig:angfit} for the ground state and the new state.
In order to determine optical model parameters for DWBA calculations, the differential cross section for elastic scattering of $^{30}$Mg on tritons and protons has been fitted using SFRESCO~\cite{thompson88}. The fitted optical potential has a deeper real part and a more shallow imaginary part than the  global optical potential. The depth of the neutron binding potential was adjusted to reproduce the binding energy. The cross section for the transfer reaction was normalized to the elastic scattering data. The theoretical differential cross section has been calculated with FRESCO~\cite{thompson88}. The choice of the optical model parameters has little influence on the shape of the angular distribution (see Fig.~\ref{fig:angfit} a)). The transfer mechanism is described solely by a one-step correlated pair transfer, since sequential two-step transfer is strongly suppressed due to the large negative $Q$-value ($-3.9$~MeV) of the t($^{30}$Mg,d)$^{31}$Mg reaction. For both states the protons clearly show an angular distribution of a $\Delta L =0$ transfer. $\Delta L=1$ and $\Delta L=2$ angular distributions clearly do not describe the data (Fig.~\ref{fig:angfit} b)).

Thus we have clearly observed a new excited $0^+$ state at $1058(2)$~keV in $^{32}$Mg. From the $\approx300$~counts in the proton spectrum for the population of the $1058(2)$~keV state one expects in case of a prompt $\gamma$-decay of this state $18(3)$ $\gamma$-proton coincidences for the 886~keV $\gamma$-ray transition. The much lower number of observed $\gamma$-proton coincidences can only have two possible causes. Either the missing part of the decay proceeds by an E0 decay directly to the ground state or the $\gamma$-decay of the nucleus occurs at a significant distance behind the MINIBALL array, leading to a reduced detection efficiency. One can easily estimate that an E0 decay branch would only be able to make a significant contribution for unphysical large $\rho^2(\text{E0})$ values and thus only a long lifetime of the excited $0^+$ state can explain the observed $\gamma$-ray spectrum. Its lifetime can be estimated, based on a GEANT4 simulation, to be larger than 10~ns. This lower limit of the lifetime corresponds to an upper limit of $B(\text{E2};\,0^+_2 \rightarrow 2^+_1) < 544$ e$^2$fm$^4$ and thus it is impossible to determine the mixing between the two $0^+$ states as any strength results in a $B(\text{E2})$ value smaller than this upper bound.

The fact that the energy of the  $0^+_2$ state is significantly lower than theoretically predicted and that both $0^+$ states are populated with comparable cross-sections potentially points to a larger mixing between the coexisting states as compared to $^{30}$Mg. In this context, it is interesting to discuss the observed cross sections in the context of their expected single-particle structure.

Monte Carlo shell-model (MCSM) calculations using the SDPF-M effective interaction~\cite{otsuka01b} predict for the ground state of $^{32}$Mg~\cite{terry08} that the two neutrons are predominately added to the $f_{7/2}$ orbital. However, the experimental cross section ($\sigma_\text{gs} = 10.5(7)$~mb) can only be described if a substantial $(\nu 2p_{3/2})^2$ contribution is added via a coherent sum $a(\nu 2p_{3/2})^2+b(\nu 1f_{7/2})^2$ with  spectroscopic amplitudes $a^2\geq0.51$ and $b=\sqrt{1-a^2}$, assuming that no other orbital plays a role. This is in qualitative agreement with results of a recent one neutron knockout experiment on $^{32}$Mg, where a large spectroscopic strength for the $3/2^-$ state in $^{31}$Mg was observed~\cite{terry08}. Both results indicate that the MCSM underestimates the $\nu 2p_{3/2}$ contribution to the ground state wave-function in $^{32}$Mg.

Without shape mixing the excited $0^+$ state in $^{32}$Mg should have a rather pure $sd$ configuration and its wave-function therefore would resemble that obtained for the ground state of $^{32}$Mg in a SM calculation with the USD interaction~\cite{wildenthal84}. Two neutron spectroscopic amplitudes with pure $sd$ wave-functions for the $0^+$ states in $^{30}$Mg and $^{32}$Mg have been calculated using OXBASH~\cite{oxbash}. They amount to -0.209, -0.184, and -0.808 for the $(1d_{5/2})^2$, $(2s_{1/2})^2$, and $(1d_{3/2})^2$ configurations, respectively.
The resulting transfer cross section is a factor of two lower than measured ($\sigma_\text{ex} = 6.5(5)$~mb). However, if a small $(2p_{3/2})^2$ contribution ($a=0.30(4)$) is added, the experimental data can be reproduced. Such a small $(2p_{3/2})^2$ contribution was observed for the $^{30}$Mg ground state in~\cite{terry08} as well ($C^2S = 0.19(7)$).

In summary, we have for the first time observed the shape coexisting excited $0^+_2$ state in $^{32}$Mg at 1058(2)~keV using an inverse kinematics (t,p) reaction of a $^{30}$Mg beam provided by REX-ISOLDE on a tritium loaded titanium foil and using the MINIBALL plus T-REX set-up. The measured angular distributions clearly identify the observed state as $0^+$ and the low $\gamma$-ray yield observed from this state suggests a lifetime of more than 10~ns. The measured cross section for the excited $0^+$ state can be described by DWBA calculations using spectroscopic amplitudes based on almost pure $sd$ configurations with a small addition of a $(p_{3/2})^2$ contribution. The current data is also consistent with previous evidence that MCSM calculations with the SDPF-M effective interaction underestimate the $\nu p_{3/2}$ component in the $^{32}$Mg ground state. This larger $(2p_{3/2})^2$ content of the wave-functions may be related to a substantial shape mixing. The lower than predicted energy of this $0^+_2$ state also poses a challenge to the theoretical description of the shape transition from $^{30}$Mg to $^{32}$Mg. In order to determine the amount of configuration mixing between the two $0^+$ states in $^{32}$Mg it will be crucial to measure the lifetime and the electric monopole strength $\rho^2(\text{E}0)$ for the $0^+_2$ state.

\begin{acknowledgments}
This work was supported by the BMBF under contracts 06MT238, 06MT9156, 06KY9136I, 06DA9036I, 06DA9041I, by the DFG Cluster of Excellence {\it Origin and Structure of the Universe}, by the EC within the FP6 through I3-EURONS (contract no. RII3-CT-2004-506065), by FWO-Vlaanderen, GOA/2004/03 and IAP P6/23 (Belgium), HIC for FAIR and US-DOE under contract number DE-AC02-05CH11231.
We would like to thank A. Dorsival and D. Carminati (ISOLDE/CERN) for the help with the target.
\end{acknowledgments}

\bibliography{draft}% Produces the bibliography via BibTeX.

\end{document}